\documentclass[aps,prb,reprint,groupedaddress,showpacs]{revtex4-1}
\usepackage{amsmath}
\usepackage{graphicx}
\usepackage{longtable}
\begin{document}

\title{Determination of ground-state and low-energy structures of perovskite superlattices from first principles}


\author{Yuanjun Zhou}
\affiliation{Department of Physics \& Astronomy, Rutgers University, Piscataway, New Jersey 08854}
\author{Karin M. Rabe}
\affiliation{Department of Physics \& Astronomy, Rutgers University, Piscataway, New Jersey 08854}

\date{\today}

\begin{abstract}
In the development of first-principles high-throughput searches for materials with desirable functional properties, there is a clear need for an efficient method to determine the ground state and low-energy alternative structures of superlattices. A method based on a simple strategy -- to generate starting structures based on low-energy structures of the constituent compounds, which are then optimized via structural relaxation calculations -- is proposed. This ``stacking method" is demonstrated on the 2:2 PbTiO$_3$/SrTiO$_3$ superlattice, which has been the subject of recent experimental and theoretical interest. Considerations relevant to wider use of the method are discussed.  
\end{abstract}



\maketitle
\section{\label{intro}Introduction}
Discovery of new functional materials with enhanced performance, novel functionalities and reduced cost and toxicity is a central goal of materials science. 
Recently, there has been tremendous progress in the synthesis of superlattices, which are artificially structured materials built up from unit-cell-scale layers of different constituent compounds\cite{Schlom1999,Schlom2001,Norton2004}. In many cases, superlattices have distinctive functional properties, which can be attributed to the strain in the layers and the high density of interfaces\cite{Neaton-apl-2003,HYHwang,Eric}.
The design of functional superlattice materials requires exploration of an enormous parameter space of constituent materials and layer sequences. 
With the development of computational techniques and resources, specifically high-throughput first-principles approaches, this process can be greatly accelerated\cite{MGI}. 

For the first-principles calculation of physical properties of a given system, determination of the ground-state structure (GSS) is the essential first step. 
Methods to predict the GSS from a given stoichiometry have recently been much discussed\cite{woodley}. 
These include genetic algorithms\cite{EAintro,bush1995,zurekEA}, random search methods\cite{pickard-rand} and data mining of experimentally-determined structural information\cite{fischer,hybrid}. While highly effective at identifying novel structures, such methods are very computationally demanding. 
Fortunately, for superlattices, the space of structures to be considered is already constrained by physical considerations, making these powerful but costly methods unnecessary. For (001) perovskite superlattices, the structures are expected to be closely related to the high-symmetry $P4/mmm$ structure generated from layered cation ordering in the ideal perovskite structure. As for pure perovskites, this high symmetry structure is expected to be unstable with respect to lower-symmetry structures with distortions such as polar distortions, oxygen octahedron rotations, and Jahn-Teller distortions. These instabilities can be identified by first-principles calculations of the phonon dispersion of the $P4/mmm$ structure, and different instabilities or combinations of instabilities can in general be expected to lead to a variety of low-energy metastable structures in addition to the ground state\cite{Rabe-springer}.

For perovskite superlattices, most first-principles studies have utilized one of three basic strategies for ground-state structure determination.
One approach focuses on instabilities identified by first-principles phonon-dispersion calculations. 
Starting with the high-symmetry reference structure, the phonon dispersion is computed, unstable modes at high-symmetry points are identified, and a set of low-symmetry structures is obtained by freezing in selected modes and relaxing the structure.
For each low-symmetry structure thus obtained, the phonon dispersion is calculated; the process terminates when the structure is at a local minimum of the energy.
Given the computational demands of phonon-dispersion calculations, this method is expensive even in the simple case of a pure perovskite; for superlattices, with larger unit cells and more modes at each wavevector, it becomes prohibitive.

In the second method, closely related to the first, the phonon dispersion is computed only for the high-symmetry reference structure.
The unstable phonon modes at high symmetry k-points are identified, and structures generated by freezing in the unstable modes, singly and in combination, are relaxed and compared, the one with lowest total energy being the GSS. A final phonon-dispersion calculation is performed to verify that the candidate GSS is stable. This method has been widely used in first-principles studies of the epitaxial-strain-induced phases of pure perovskites \cite{Oswaldo,JunheeSrMnO3} and ultrashort period (1:1) superlattices of perovskites\cite{Eric,Eric-thesis, Javier1, Yuanjun}.

A quite different strategy is to generate starting structures by making small random displacements of the atoms away from the high-symmetry $P4/mmm$ structure, relax each starting structure and compare the distinct structures thus generated.
This has the advantage of sampling the relevant structure space without any particular bias, but it is relatively demanding, as 
a general random initial configuration will take a large number of iterations to converge to the nearest minimum.
Further, in principle this method could miss local minima or even the ground state due to statistical fluctuations, with no guarantees even if the number of starting structures is systematically increased.
Therefore, this method is best used as a complementary ``double check,'' to make sure that no exotic low-energy structures have been missed by other methods.

Given the ground state structure of the superlattice, it would be implausible if the structure of an individual layer were to derive from a high-energy bulk structure. Indeed, it has previously been assumed that the structures of each constituent layer should derive from the ground state of the corresponding pure compound at the relevant epitaxial strain.
If a starting structure is obtained by distorting each layer to its ground state structure, and then relaxed, the expectation is that the original distortions would remain, and additional distortions of certain types (specifically out-of-plane polarization and oxygen octahedron rotations about an in-plane axis) in one layer would induce the same distortion in an adjacent layer through considerations of electrostatic boundary conditions and steric constraints associated with the rigidity of oxygen octahedra.
However, it should be noted that the structure of an individual layer could be derived not from the ground state, but from a distinct low-energy alternative state of the corresponding pure compound at the relevant epitaxial strain. In that case, the ``bulk" energy cost would be more than balanced by a reduction in energy associated with matching conditions or interface energetics.
Thus, the assumption above should be modified to be that structures of the constituent layers will derive from a low-energy state of the pure compound at the relevant epitaxial strain, and that the ground state and low-energy states of the superlattice can be obtained by relaxing starting structures obtained from stacking combinations of the low-energy pure-compound states.

In this paper, we propose a simple and efficient stacking method, suitable for high-throughput studies, to determine the GSS and low-energy structures in perovskite superlattices based on this modified assumption.
In section \ref{method}, we describe the method in detail. 
In section \ref{results}, we demonstrate the method by application to the structure determination of the 
(PbTiO$_3$)$_2$(SrTiO$_3$)$_2$ superlattice as a function of epitaxial strain, as previous work suggests that this system is particularly rich in low-energy structures\cite{Eric,Javier1}. 
We first describe the construction of a database of the low energy structures of the constituent pure compounds PbTiO$_3$ and SrTiO$_3$. 
We then describe the results of structure determination of (PbTiO$_3$)$_2$(SrTiO$_3$)$_2$ via the stacking method, illuminating various aspects of implementation of the method. At 0\% strain, we find that (PbTiO$_3$)$_2$(SrTiO$_3$)$_2$ does not have a single GSS, as previously proposed\cite{Javier1}, but has a flat GS energy landscape. This has important implications for experimental studies of the structure and properties of this system.

\section{\label{method}Methods}

 \begin{figure}
 \includegraphics[width=0.5\textwidth]{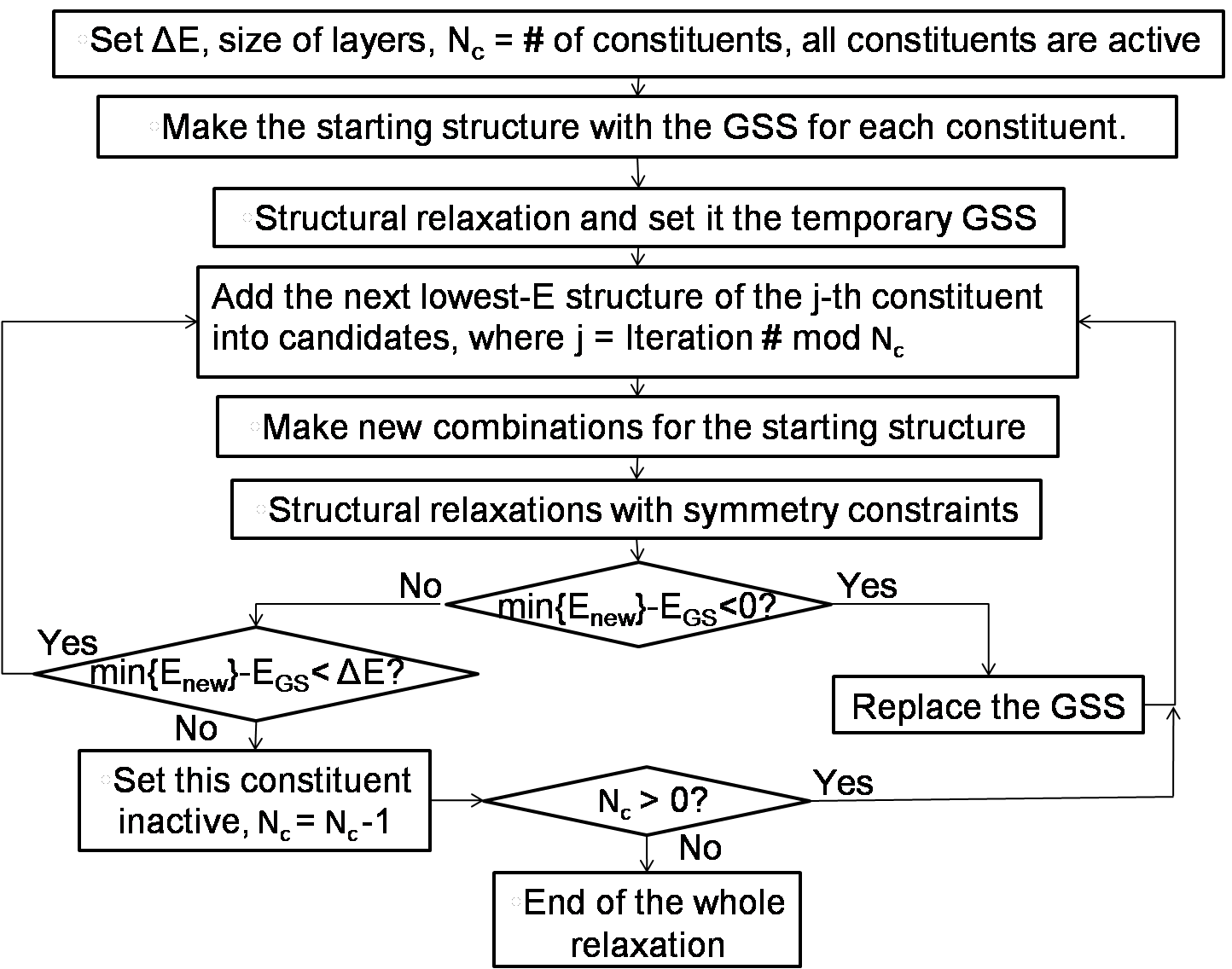}%
 \caption{\label{flow}Flowchart for the stacking method for identification of the ground-state structure of a superlattice.}
 \end{figure}

Our stucture determination method, which we refer to the ``stacking method," is based on the assumption that for the GSS of the superlattice, structures of the constituent layers will derive from a low-energy state of the pure compound at the relevant epitaxial strain.
Based on this assumption, we construct starting structures by putting each constituent layer into a low-energy pure-compound structure, generating all such symmetry-inequivalent combinations.
Relaxations from these starting structures using Hellmann-Feynman forces and stresses are performed to minimize the overall energy of the superlattice structure. This energy of the superlattice includes the energy associated with electrostatic interactions of the layers, steric constraints associated with rigidity of the oxygen octahedra, and contributions from the interface regions.
 
Our method is progressive, starting with combinations of the ground states of the constituents and adding low-energy constituent structures into the combinations until no more low-energy structures for the superlattice are found. For this, we choose an energy window: when the process yields superlattice structure energies that are above the window, the process terminates. Thus, in addition to the GSS, our method will identify the lowest-energy alternative structures as well, which can be of interest for functional properties.

Once the constituents for the superlattice are chosen, the first step is to generate the database of low-energy states of the pure constituent compounds at the relevant epitaxial strain. This can be done using a conventional method such as the second method mentioned in Section \ref{intro}. In a study including various combinations of several different constituents, it is convenient to generate the database for all constituents as a pre-processing step.  

The steps in the stacking method for structure determination of a superlattice with specified layer thickness then follow the flow chart shown in Fig. \ref{flow}:

(1) We set the energy window $\Delta E$. We define $N_c$ to be the number of active constituents.
If a constituent has been given ``inactive'' status, we will not include any additional higher-energy structures for the relevant constituent. At the beginning, all constituents are active.

(2) We use the lowest energy structure for each constituent to construct the starting structure for the superlattice, with the atomic positions of the interfacial layers being the linear combination of the two adjacent constituents.

(3) We relax the structure and set it as the candidate GSS.

(4) Considering each constituent in turn, we take the next lowest-energy structure of the given constituent and construct ``new" starting structures by combining it with the low-energy structures for the other constituents already included. 

(5) We do structural relaxations on the new starting structures.

(6) If the minimum of the energy of the ``new'' superlattices, min\{$E_{new}$\}, is lower than the current ground state energy $E_{GS}$, then we replace the GSS with the ``new'' superlattice of the lowest energy, and then return to step (4). 

(7) If $0 <$ min$\{E_{new} \}-E_{GS} < \Delta E$, we check to see if any of the ``new" structures in the energy window are distinct structures not already identified.  We add these to the list of low-energy structures and return to step (4). If no new low-energy structures are found within the energy window, then we are done with adding low-energy structures for this constituent. We declare this constituent to be inactive and decrease the number of active constituents $N_c$ by one. If there are still active constituents, we go to step (4). 

(8) If only the ground-state structure is desired, the process now terminates, with the identification of the ground state structure as the lowest energy structure found. If the identification of low-energy alternative structures is also desired, there is one additional step (not included in the ground-state search flowchart in Fig. \ref{flow}). For each low-energy superlattice structure $S$ already identified, the full distortion patterns in each constituent layer are analyzed. All symmetry-inequivalent combinations of the distorted layer structures (those with reversal of the out-of-plane polarizations or in-plane rotations can be excluded as being much higher in energy) are generated. Those that are not symmetry-equivalent to $S$ are included as additional starting structures and relaxed. This step will be explained in more detail in the discussion of the application to (PbTiO$_3$)$_2$(SrTiO$_3$)$_2$ superlattice.

Structural relaxations in this approach preserve space group symmetries, so that the space group of the relaxed structure will be the same or a supergroup of the space group of the starting structure. In the latter case, the iterative relaxation process will in general yield a structure that has tiny displacements of the atoms that break the symmetry of the supergroup.
For example, if we start with a low-symmetry configuration with polar $P4mm$ which relaxes to a nonpolar $P4/mmm$ structure, the relaxed structure would in general have tiny displacements away from the $P4/mmm$ structure resulting in a space group of $P4/mmm$. 
For this reason, most space-group-identification software tools find the highest-symmetry space group consistent with displacements of the atoms by a specified distance, referred to as the tolerance.
If the displacements in the case of the relaxed structure in the example above are less than the tolerance, the space group will be identified as $P4/mmm$.
In the first-principles calculation, a very low tolerance ($10^{-5}\AA$) is chosen to avoid artificially increasing the symmetry during the calculation.
Using the python package ``pyspglib'' \cite{pyspglib}, we analyze the relaxed structure by increasing the tolerance from $10^{-5}\AA$ until we find the critical tolerance (CT) at which the space group changes from the space group of the starting structure to one of its supergroups. 
A small CT suggests that the structure has relaxed into a structure with a higher symmetry space group. 
The upper limit on CT which establishes relaxation to the supergroup is $CT_{UL}=2\delta E/\delta F$, where $\delta F$ is the force threshold in the relaxation and $\delta E$ is the energy resolution. 
We will discuss explicitly how the the upper limit on CT is chosen for the example of 2:2 in the next section.

Our calculations for PTO/STO were performed  using the local density approximation\cite{LDA1,LDA2} implemented in the {\it Vienna Ab initio Simulation Package} (VASP-5.2.12)\cite{vasp1,vasp2}.
The projector augmented wave (PAW) potentials\cite{paw} used contain 10 valence electrons for Sr ($4s^24p^65s^2$), 14 for Pb ($5d^{10}6s^26p^2$), 10 for Ti ($3p^63d^24s^2$).
We used a 500 eV energy cutoff, $\sqrt 2\times\sqrt 2\times 4$  supercell and $4\times 4\times 1$ Monkhorst-Pack(MP) k-meshes\cite{MP} for total energy calculations in structural relaxation, and the force threshold is $\delta F = 5\times 10^{-3}$ eV/$\AA$. The energy resolution is 1 meV/5atom, so that $\delta E = 0.2$ meV/atom and the $CT_{UL}$ is thus 0.08\AA.

\section{\label{results}Results}
\subsection{\label{preprocess}Low energy structures of pure compounds: PbTiO$_3$ and SrTiO$_3$}
 \begin{figure}
 \includegraphics[width=0.5\textwidth]{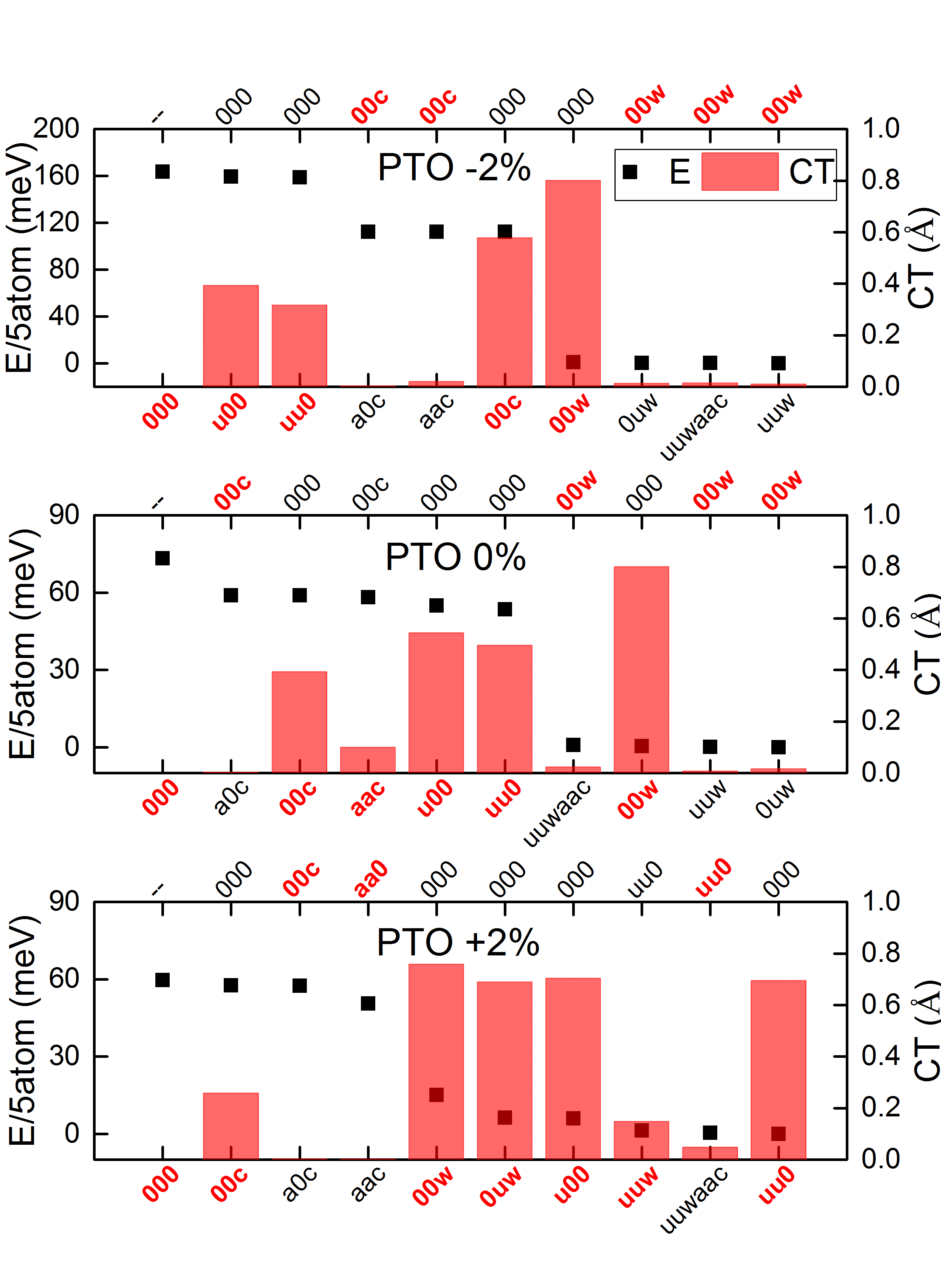}%
 \caption{\label{PTOmerge}Total energies (black squares) and space-group-symmetry analysis for relaxed structures of epitaxially-strained PTO. Top, -2\% strain. Middle, 0\% strain. Bottom, +2\% strain. Energies are in meV per 5 atoms, with the zero of energy for each strain taken as the energy of the ground state structure at that strain. The horizontal axis is labeled at the bottom by the space group of the starting structure, and at the top by the space group of the supergroup produced at values of the tolerance higher than the critical value CT, which is shown as a red bar.
Stable distortions are typeset in bold red.}
 \end{figure}

\begin{figure}
 \includegraphics[width=0.5\textwidth]{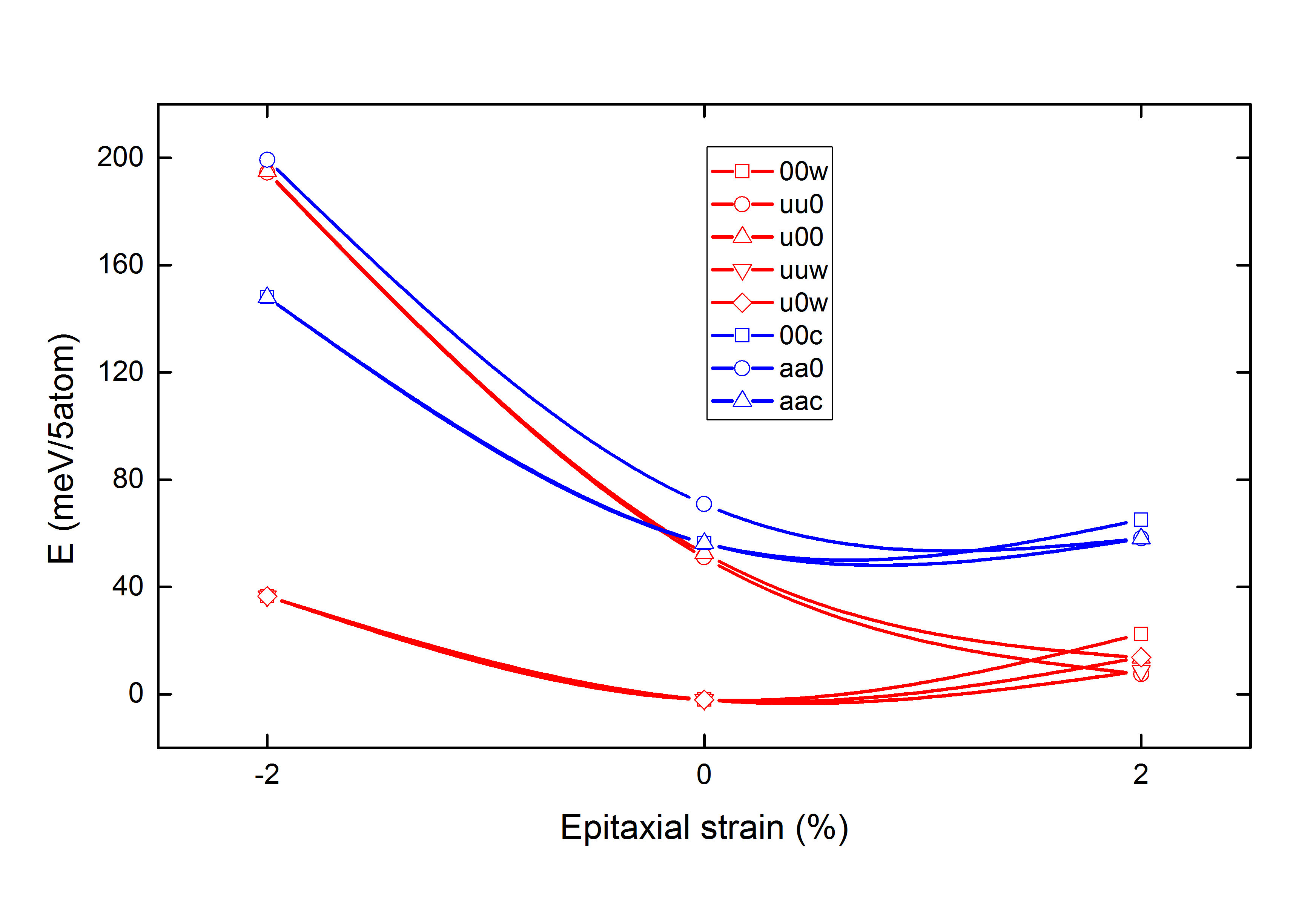}%
 \caption{\label{3ptPTO}Energies of low-energy distorted structures of PTO as functions of epitaxial strain. Red lines represent structures with polar distortions, and blue lines represent structures with octahedron rotations. }
 \end{figure}

 \begin{figure}
 \includegraphics[width=0.5\textwidth]{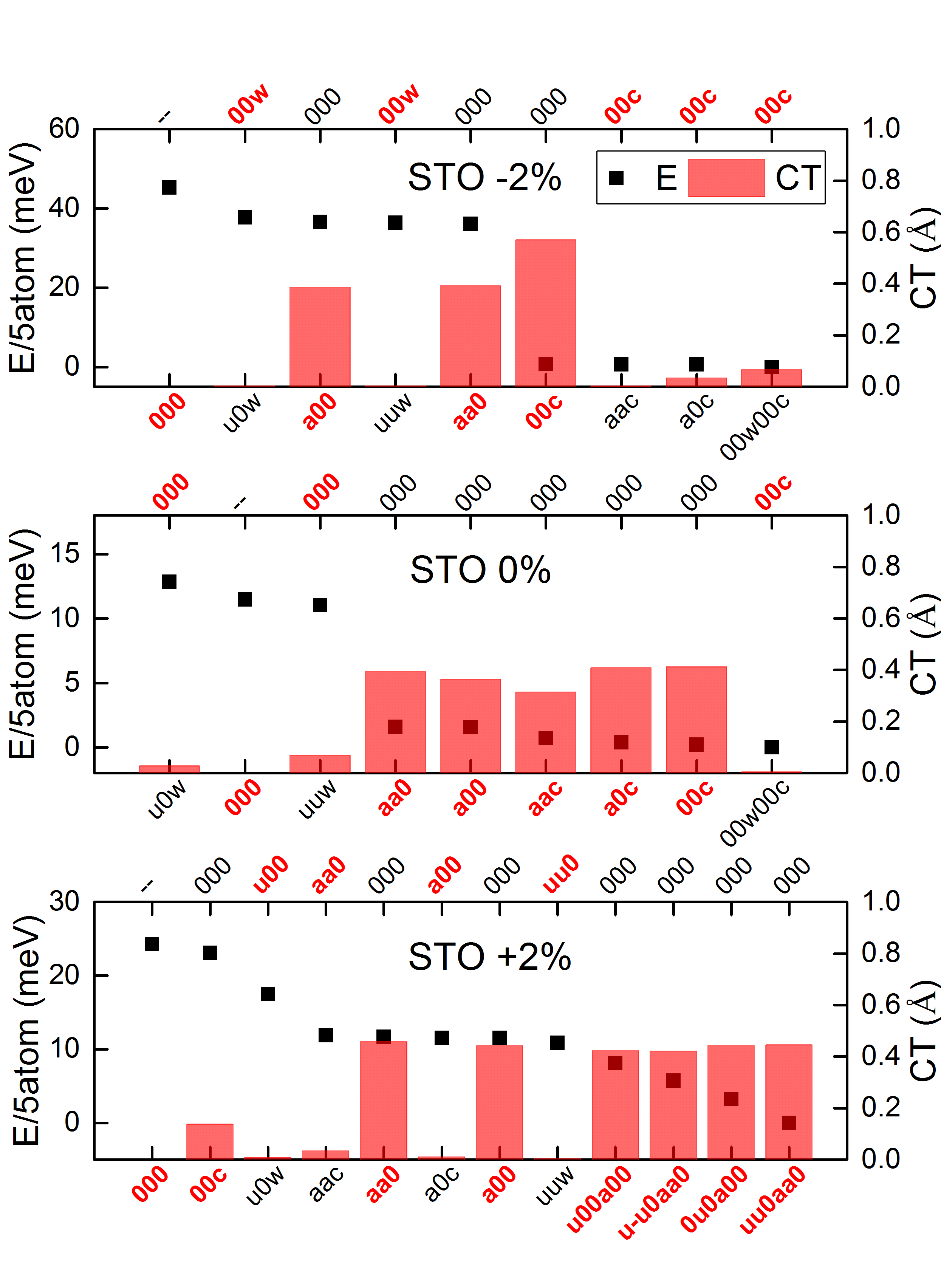}%
 \caption{\label{STOmerge}Total energies and space-group-symmetry analysis for relaxed structures of epitaxially-strained STO. Conventions as in Figure \ref{PTOmerge}.}
 \end{figure}

\begin{figure}
 \includegraphics[width=0.5\textwidth]{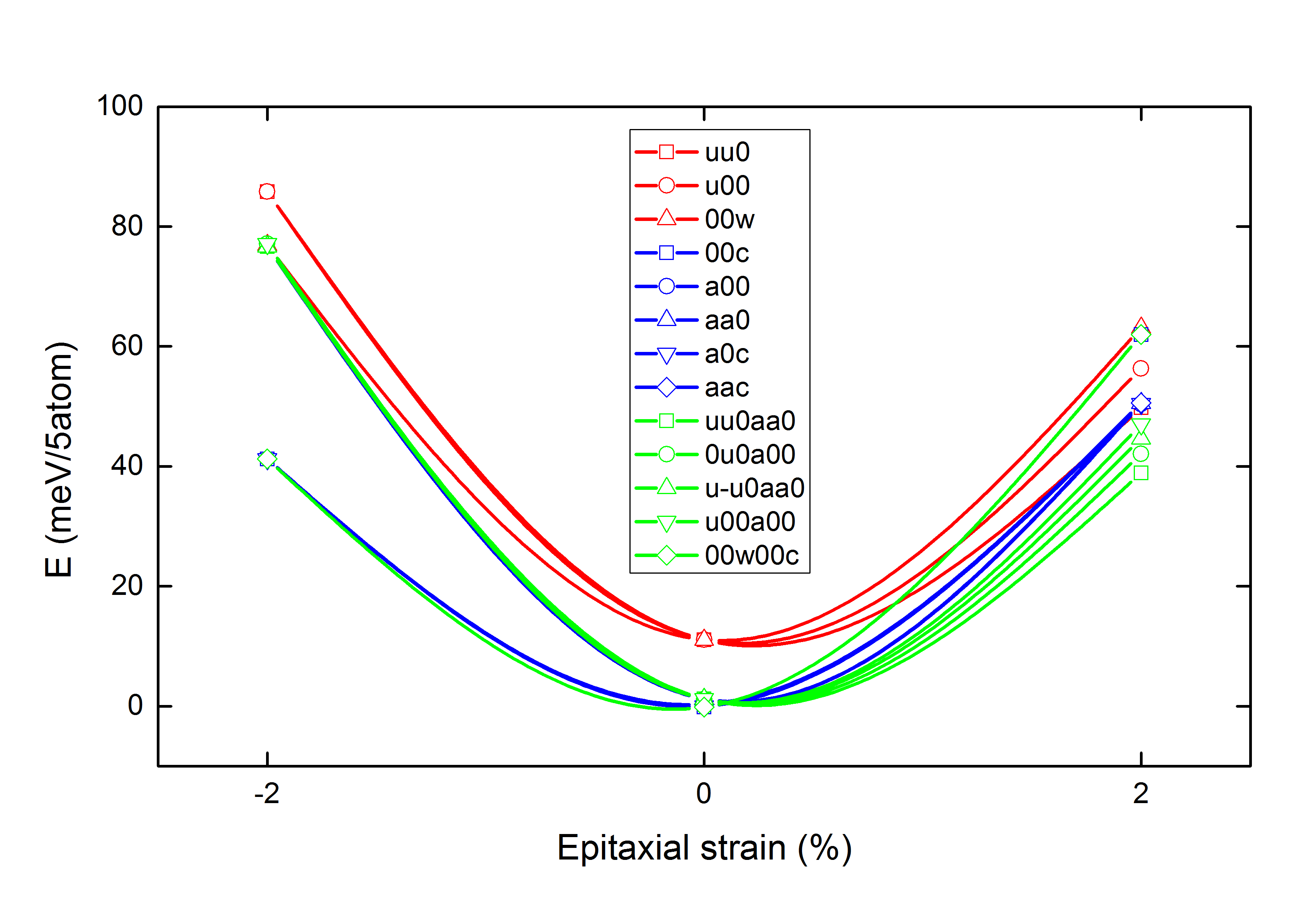}%
 \caption{\label{3ptSTO}Energies of low-energy distorted structures of STO as functions of epitaxial strain. Red lines represent structures with polar distortions,  blue lines represent structures with octahedron rotations, and green lines represent structures with combined distortions of polar modes and rotations.}
 \end{figure}

In the stacking method, the first step in structure determination of a superlattice is construction of the database of the low energy structures of the constituent pure compounds at the relevant epitaxial strain(s). 
For our demonstration case (PbTiO$_3$)$_2$(SrTiO$_3$)$_2$, the constituents are PTO and STO. 
First-principles computations of the phonon dispersions for the cubic-perovskite high-symmetry reference structures show that in both compounds, $\Gamma_{15}$ and $R_{25}$ modes are unstable\cite{PTOphonon,STOphonon}. 
To identify the low-energy structures for each compound at each epitaxial strain, we freeze in the unstable $\Gamma$ and R modes, singly and in combinations, and relax the structures (we note that not all structures thus obtained will be distinct). The space group of each relaxed structure is identified by using the CT approach with a threshold of 0.08\AA. Here we label the structures not by the space group, but by the distortions that generate them, indicated in Cartesian components. $u$ and $w$ denote nonzero in-plane and out-of-plane polar mode components, respectively, while $a$, $c$ denote non-zero R-point octahedron rotations around in-plane and out-of-plane axes, respectively.

Figs. \ref{PTOmerge} and \ref{3ptPTO} show the relaxed total energies and CTs for bulk PTO for various combinations of polar distortions and R point octahedron rotations at three values of epitaxial strain (-2\%, 0\% and +2\%, defined with respect to 3.849\AA, the computed lattice constant of cubic STO).
For -2\% and 0\% epitaxial strain, the most favorable distortion is the out-of-plane polar distortion, denoted by $00w$, while for +2\%, the states with in-plane polarization and out-of-plane polarization are essentially equal in energy, consistent with previous studies\cite{PTO-thermo, PTO-landau,STO-oswaldo,Bungaro-PTO1,Bungaro-PTO2}. 
The low-energy structures at each strain, in order of increasing total energy, are listed in Table \ref{preprocess} and constitute the required database for PTO. 

Figs. \ref{STOmerge} and \ref{3ptSTO} show the analogous results for STO. 
For -2\% epitaxial strain, the out-of-plane oxygen-octahedron rotation $00c$ is most favorable.
For +2\% epitaxial strain, in-plane polar distortions combined with in-plane octahedron rotations produce the lowest energy structures, with different combinations of these two distortions resulting in slightly different energies, in agreement with previous studies\cite{STO-strain,STO-Antons,STO-Li,STO-oswaldo}.  
The 0\% epitaxial strain case is the most complicated: five distinct structures with different rotation patterns have almost identical energy. This suggests a rather flat energy surface. 
As for PTO above, the database of low-energy structures for STO is given in Table \ref{preprocess}.

\begin{table}
 \caption{\label{preprocess}The low energy distorted structures of PTO and STO for -2\%, 0\% and +2\% strain.}
 \begin{ruledtabular}
 \begin{tabular}{ll}
{\bf Perovskite} & {\bf Stable distortions} \\
PTO -2\% & $00w$, $00c$,  $uu0$,$u00$\\
PTO 0\% & $00w$, $uu0$, $u00$, $aac$, $00c$ \\
PTO +2\% & $uu0$, $uuw$, $u00$, $u0w$, $00w$, $aa0$, $00c$\\
STO -2\% &  $00c$, $aa0$, $00w$,$a00$\\
STO 0\% & $00c$, $a0c$, $aac$, $a00$, $aa0$\\
STO +2\% & $uu0aa0$, $0u0a00$, $u$-$u0aa0$, $u00a00$,\\
& $uu0$, $a00$, $aa0$, $u00$, $00c$\\

 \end{tabular}
 \end{ruledtabular}
 \end{table}

\subsection{\label{stru_deter}Structure determination for (PTO)$_2$/(STO)$_2$}

\begin{table}
 \caption{\label{PTOSTO-2} Space groups of (PTO)$_2$/(STO)$_2$ starting structures for -2\% strain.
The space group information is obtained using ISOTROPY\cite{ISOTROPY,isotropy-link}.
The labels \#1 and \#2 differentiate between two inequivalent starting structures with the same space group.}
 \begin{ruledtabular}
 \begin{tabular}{ccccc}
PTO/STO	&	$00c$	&	$aa0$	&	$00w$	& 	$a00$	\\
$00w$	&	$P4bm\#1$	&	$Pma2$	&$P4mm$	& $Cmm2$	\\
$00c$	&	$P4/mbm$	&	$P2_1/c$ 	&$P4bm\#2$	    	&	$C2/m$	\\
 \end{tabular}
 \end{ruledtabular}
 \end{table}

\begin{table}
 \caption{\label{PTOSTO+2}  Space groups of (PTO)$_2$/(STO)$_2$ starting structures for +2\% strain. Conventions as in Table \ref{PTOSTO-2}. The triplets in parenthesis indicate the origin, if there are two or more structures detected in the same space group yet different coordinates.}
 \begin{ruledtabular}
 \begin{tabular}{ccccc}
PTO/STO	&	$uu0aa0$	&$0u0a00$		&	$u$-$u0aa0$	& $u00a00$		\\
$uu0$	&	$Pnc2$	&	$Pc\#1$ 	& $Pc\#2$	&$Pc\#3$	\\
             &	&$(00{1\over 4})$ 	&($00{1\over 4}$)	&	($00{1\over 4}$)\\
$u$-$u0$ &	$Pc\#4$	&		$Pc\#5$ &$Pmn2_1$    	&	$Pc\#6$	\\
          & ($00{1\over 4}$)          & $(00{1\over 4})$  &       & $(00{1\over 4})$ \\
$uuw$	&	$Pc\#7$	&	$P1\#1$	  	&	$P1\#2$    	&	$P1\#3$\\
 & $({1\over 4}00)$& & & \\
$u$-$uw$ &	$P1\#4$ 	&$P1\#5$ &	$Pm$ &  $P1\#6$	\\
$u00$ &	$Pc\#8$	&$Pc\#9$	 	&	$Pc\#10$    	&	$Abm2\#1$	\\
          & ($00{1\over 4}$)    & $(00{1\over 4})$       &$(00{1\over 4})$        & \\
$0u0$ & $Pc\#11$ &$Pc\#12$ & $Pc\#13$&  $Abm2\#2$\\
         & ($00{1\over 4}$)          & $(00{1\over 4})$  & $(00{1\over 4})$     &  $(00{1\over 4})$\\
$u0w$ &	$P1\#7$	&	$P1\#8$	   	&$P1\#9$&	$Cm$\#1 	\\
 & &  & & (000)\\
$0uw$ & $P1\#10$&$Cm\#2$  &$P1\#11$ &$P1\#12$  \\
 & &$({1\over 2}00)$ & &  \\

 \end{tabular}
 \end{ruledtabular}
 \end{table}

\begin{table}
 \caption{\label{PTOSTO0} Space groups of (PTO)$_2$/(STO)$_2$ starting structures for 0\% strain. Conventions as in Table \ref{PTOSTO-2} and \ref{PTOSTO+2}. The dash indicates that the structure is equivalent to the one above it.}
 \begin{ruledtabular}
 \begin{tabular}{cccccc}
PTO/STO	&	$00c$	&	$a0c$	& 	$aac$	&	$a00$	&	$aa0$	\\
$00w$	&	$P4bm$	&	$Cm$\#1	&	$Pc$\#1	&  	$Cmm2$	&	$Pma2$	\\
	&	 	&	($1\over 2$00)	& 	($1\over 4$00) 	&	 	&	 	\\
$uu0$	&	$Pmc2_1$ 	&	$P1\#1$	&	$Pc\#3$ 	&	$Pc\#2$ 	 &	$Pnc2$    	\\
	&	  	&	 	&	(${1\over 4}00$)  	&	($00{1\over 4}$)	&	 	\\
$u$-$u0$	&	-	&	-	&	$P2_1$	&	-	&	$Pmn2_1$	\\
$u00$	&	$Amm2$\#1 	&	$Cm$\#2 	&	$P1\#2$	&	$Abm2$\#1	&	$Pc$\#4	\\
	&	($1\over 2$0$1\over 4$)	&	($1\over 2$00)	&		&	(00-${1\over 4}$)	&	($00{1\over 4}$)	\\
$0u0$	&	-	&	$C2$	&	-	&	$Abm2$\#2	&	-	\\
	&	 	&	 	&	 	&	(-${1\over 4}$-${1\over 4}$-${1\over 4}$)	&	 	 	\\
 \end{tabular}
 \end{ruledtabular}
 \end{table}

In this section, we describe the starting structures and the
ground-state and low-energy structures of (PTO)$_2$/(STO)$_2$ 
found through our structural determination procedure with an energy window of 30 meV/5 atoms for -2\% epitaxial strain and 15 meV/5 atoms for 0\% and +2\% epitaxial strain.
As discussed in the previous section, in both compaounds, at -2\% there is a single structure much lower in energy than the others, and it is expected that the GSS of the superlattice will be a stacking of these two structures. At 0\% and +2\% strain, the near-degeneracy of several low-energy states in one or both compounds is expected to lead to a less clear-cut situation requiring a systematic approach; it is here that our stacking method will yield nontrivial results.

\begin{figure}
 \includegraphics[width=0.5\textwidth]{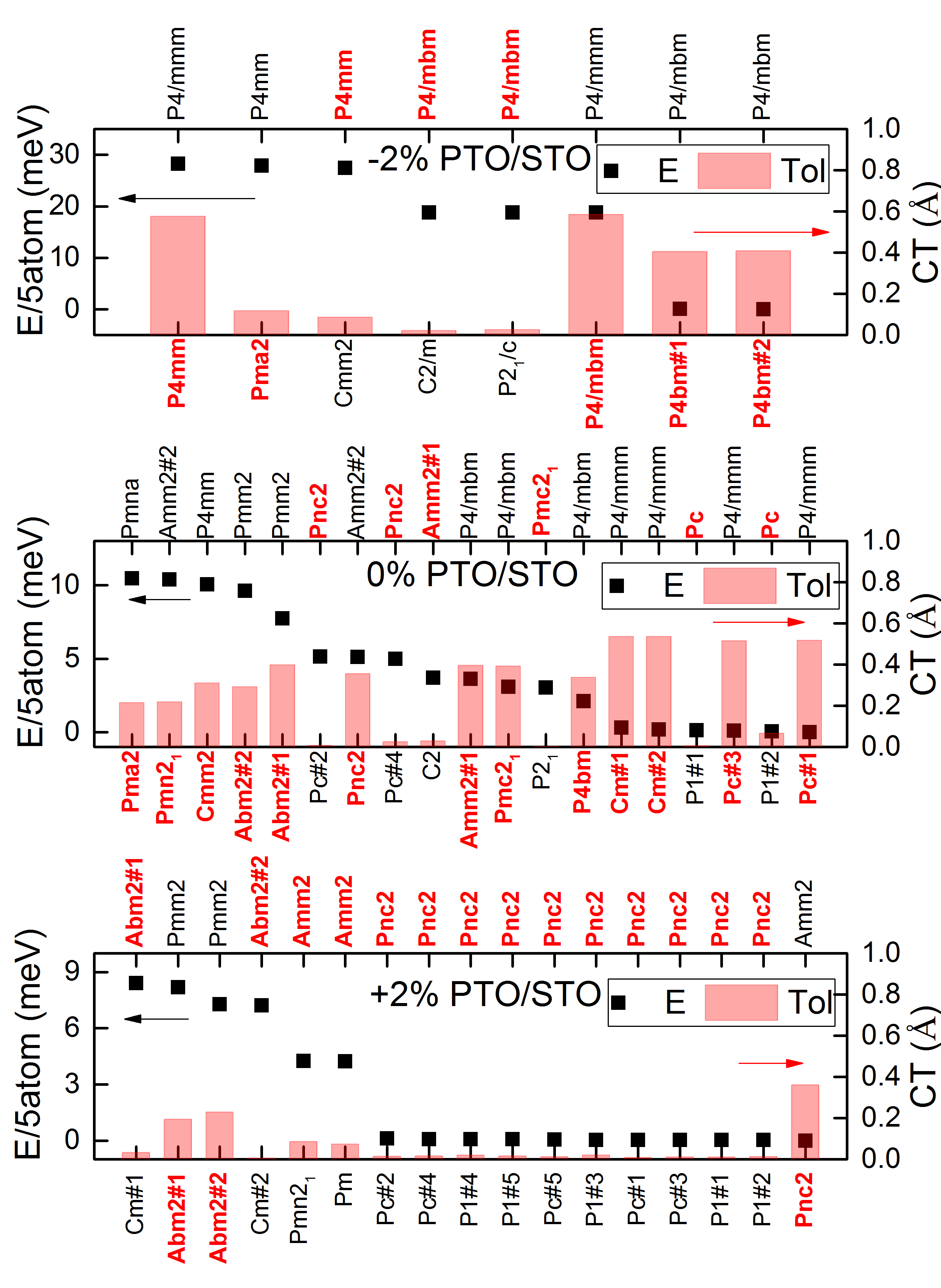}%
 \caption{\label{PTOSTOdata}Total energies and space-group-symmetry analysis for relaxed structures of the epitaxially-strained 2:2 PTO/STO superlattice. Conventions as in Figure \ref{PTOmerge}.}
 \end{figure}

The simplest case is for -2\% strain. 
The iterative process terminates after only eight starting structures, shown in Table \ref{PTOSTO-2}.
These starting structures are obtained by combining the two lowest energy structures of PTO at -2\% strain with the four lowest energy structures of STO at -2\% strain (the rows and columns, respectively, of Table \ref{PTOSTO-2}).
As shown in the top panel of Fig. \ref{PTOSTOdata}, the GSS is $P4bm$, obtained from the $P4bm\#1$ starting structure built from the ground state of PTO ($00w$) and STO ($00c$), with octahedron rotations and polar displacements along [001], consistent with previous results\cite{Javier1}.  The $P4bm\#2$ starting structure also relaxes to this structure.
In this $P4bm$ state, the interlayer interactions induce octahedron rotations in the PTO layer and out-of-plane polarization in STO layer.
Rotations in TiO$_2$ layers between PbO layers and those between SrO layers are in the same direction and with surprisingly close amplitudes. Rotations in the two interfacial TiO$_2$ layers are in opposite directions, so that one is in the same direction as the bulk layers and the other opposite.
Above the GSS we find two unstable saddle point structures, $P4/mbm$ and $P4mm$ which are supergroup of the GSS $P4bm$, and a distinct structure, $Pma2$, with small amplitudes of octahedron rotations along [110]. 


For +2\% strain, there are many more starting structures than in the -2\% strain case. 
First, there are many low-energy states of STO at this strain.   
Also the low-energy states of STO and PTO include in-plane distortions along [110] or [100], as shown in Table \ref{preprocess}, and so there can be multiple symmetry-inequivalent ways to combine distortions due to different relative orientations of the in-plane distortions. 
The majority of starting structures (26 out of 32) relax to the GSS $Pnc2$ structure, with the polar distortion along [110] and octahedron rotations around [110], as shown in the bottom panel of Fig. \ref{PTOSTOdata} (the data for 15 of the $Pc$ and $P1$ starting structures are not shown; they all relax to the GSS $Pnc2$ structure). 
Octahedron rotations are induced in PTO layers due to the interlayer interactions, with amplitudes similar to those in STO layers. 
However, in contrast to the large energy difference between distinct structures for -2\%, here the energy scale for alternative low-energy states is smaller, due to the smaller energy differences for stable distortions in +2\% strained bulk STO. 
We also find a unstable saddle point structure above the GSS,
$Amm2$, which is the supergroup of the GSS $Pnc2$, 
and other alternative low-energy structures $Abm2_{low}$, $Abm2_{high}$, and $Cm$.


The most complicated case is for the intermediate value of 0\% strain. 
Low-energy states of STO and PTO both with in-plane distortions, characteristic of tensile strain, and out-of-plane distortions, characteristic of compressive strain, are represented in the starting-state combinations, given in Table \ref{PTOSTO0}.
We find that the $P4bm$ stacking of the two ground states for the constituent compounds is in fact not the ground state structure for the superlattice.
Further, the GSS is not unique: as seen in the middle panel of Fig. \ref{PTOSTOdata}, relaxation identifies two distinct structures, $Cm$ and $Pc$, with an energy difference less than 0.3 meV/5 atoms, less than the resolution of our calculation.
Both the $Pc$ and $Cm$ structures have octahedron rotations and polar displacements in each constituent layer.
The difference is that the direction of in-plane distortions is [110] for $Pc$ and is [100] for $Cm$, suggesting a ``flat'' energy surface for in-plane distortions. 
The ground-state $Pc$ structure has been previously identified and discussed\cite{Javier1}. The present results suggest that the experimental determination of the low-temperature structure would not show $Pc$ as a well-defined ground state, but that the results would show variations in the directions of the distortions resulting from the flat ground-state energy landscape which could also have an impact on the physical properties. 
In addition to the characterization of the ground state, at this stage we find a number of low-energy unstable saddle point structures. 
$P4bm$, $Pmc2_1$, $Amm2$, $Pnc2$, $Abm2$, $Cmm2$, $Pmn2_1$, $Pma2$.
They are all supergroups of the GSS $Pc$ or $Cm$ structures.

\begin{figure}
 \includegraphics[width=0.5\textwidth]{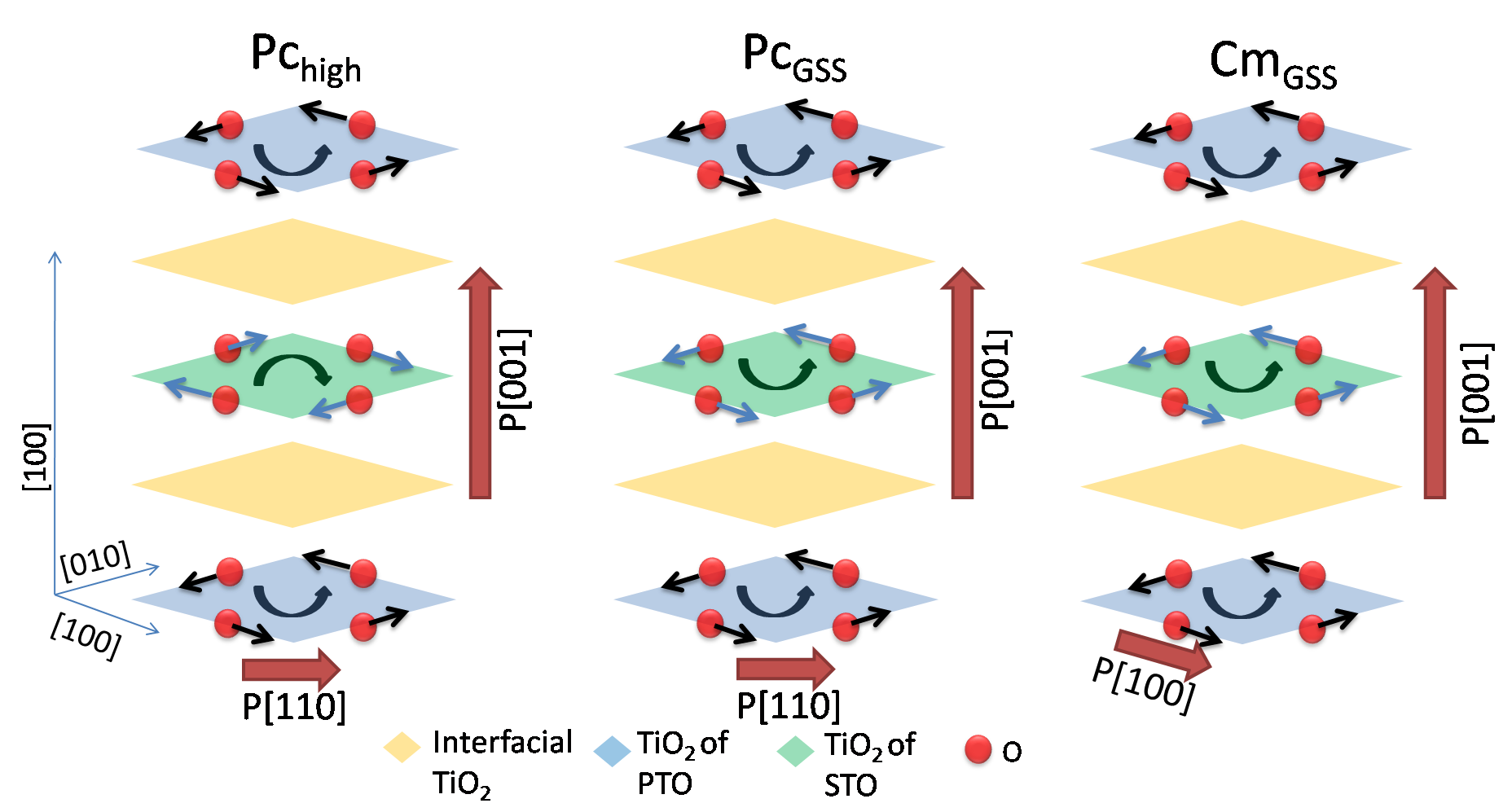}%
 \caption{\label{Pc_str}Oxygen octahedron rotation patterns and polarization directions for the $Pc_{GSS}$, $Pc_{high}$  and $Cm_{GSS}$ structures of the 2:2 PTO/STO superlattice. The rotations of the two interfacial TiO$_2$ planes (yellow) are in the same sense in all three structures and are not shown. The blue and green planes represent the TiO$_2$ layer between SrO layers and PbO layers, respectively. Note that the sense of the rotation in the central layer of $Pc_{high}$ is opposite to that of the rotation in the central layer of $Pc_{GSS}$ and $Cm_{GSS}$.}
 \end{figure}

Finally, we carried out the final step described in Sec. \ref{method} to identify additional low-energy structures. First, for 0\% strain, we analyzed the full distortion patterns in each constituent layer for the $Pc$ GSS, which can be characterized as $uuwaac$ (in-plane polarization along [110], out-of plane polarization along [001], octahedron rotation around [110] and octahedron rotation around [001]. Excluding the configurations with high energy due to electrostatics or steric constraints, these can be combined in four ways: $uuwaac$+$uuwaac$, -$u$-$uwaac$+$uuwaac$, $uuwaa$-$c$+$uuwaac$ and -$u$-$uwaa$-$c$+$uuwaac$. Two of these, $uuwaac$+$uuwaac$, -$u$-$uwaac$+$uuwaac$, relaxed to the $Pc$ GSS,
while the other two relax to a distinct $Pc$ structure with energy 2 meV/5 atoms above the ground state, which we denote as $Pc_{high}$.
The $Pc_{high}$ structure is closely related to the ground state $Pc$ structure, the main difference being the pattern of oxygen octahedron rotations around [001]. As shown in Fig. \ref{Pc_str}, for the $Pc_{high}$ state the octahedron between SrO layers rotates in the opposite sense to the one between PbO layers, while for the $Pc$ GSS state, these two octahedra rotate in the same direction.

\begin{figure}
 \includegraphics[width=0.5\textwidth]{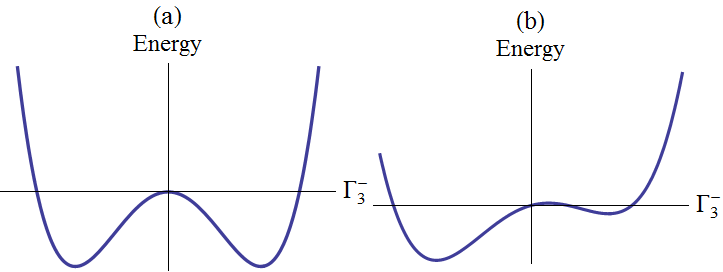}%
 \caption{\label{doublewell}Schematic curves for the total energy as a function of the amplitude of an unstable mode in two cases: trilinear terms including the mode are (a) forbidden by symmetry, or (b) allowed by symmetry.}
 \end{figure}

The reason that this low-energy structure was not identified in the earlier steps of structure determination is that the starting structures did not contain any oxygen octahedron rotation in the PTO layer (see Table \ref{PTOSTO0}). The rotation in the PTO layer in the relaxed structure is induced by the symmetry breaking for the superlattice by the rotations in the STO layer, picking out one of two senses for the rotation. 
Fig.\ref{doublewell} shows an analogous coupling of distortions in a pure perovskite, showing two inequivalent local minima obtained by freezing a $\Gamma^-_3$ mode into a structure obtained by freezing in a $M^+_3$ and a $M^-_1$ mode; symmetry analysis shows that in the energy expansion around the high-symmetry cubic structure there is a term trilinear in $M^+_3$, $M^-_1$ and $\Gamma^-_3$. By relaxation of the starting state with $M^+_3$ and $M^-_1$ nonzero and $\Gamma^-_3$=0, only the lower minimum would be found.

With the same procedure applied to the $Cm$ GSS at 0\% strain and the $P4bm$ GSS at -2\% strain, we find a $Cm_{high}$ state at 0\% with distortion pattern $u0wa0$-$c$+$u0wa0c$ and energy 2 meV/5 atoms, and a $P4bm_{high}$ state at -2\%, with distortion pattern $00w00$-$c$+$00w00c$ and energy 3 meV/5 atoms. The appearance of inequivalent local minima thus appears to be relatively common in superlattices. 
The idea that the low-energy landscape is complex for small epitaxial strain, as observed in a previous study\cite{Javier1}, is here strengthened by the fact that even more distinct structures at small scales of differences in total energies are found by the stacking method than previously recognized.

\begin{figure}
 \includegraphics[width=0.5\textwidth]{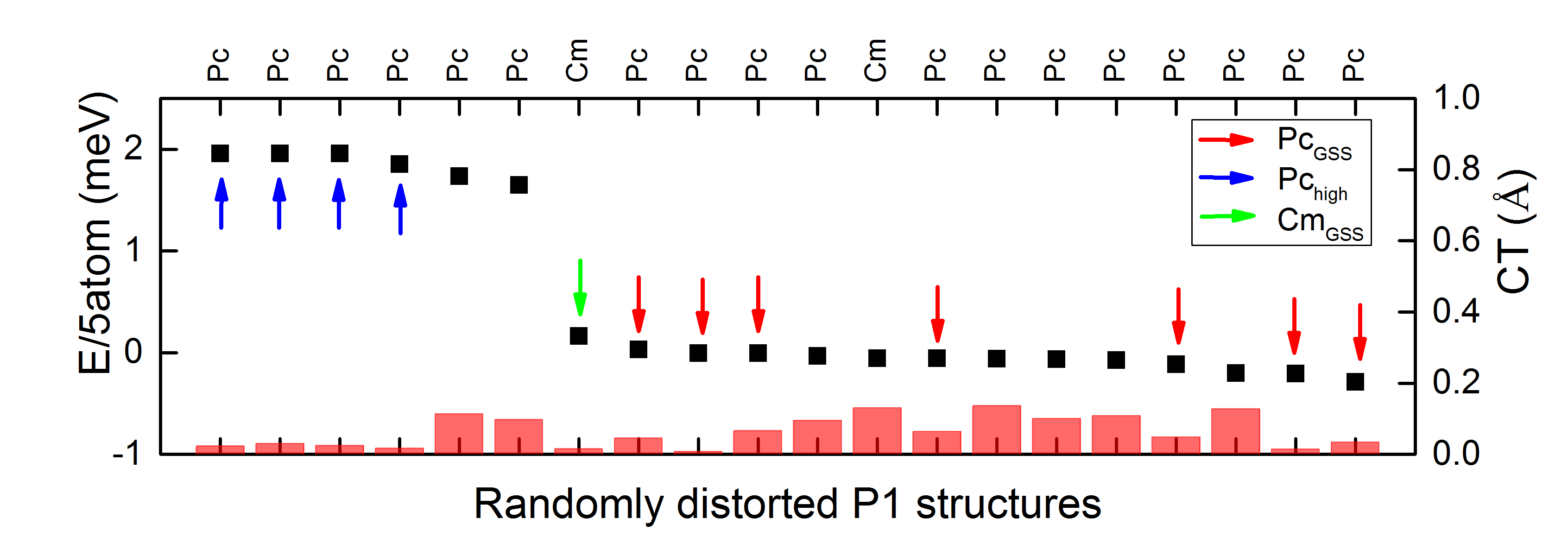}%
 \caption{\label{randomP1}Total energies and space-group-symmetry analysis for relaxed structures of the epitaxially-strained 2:2 PTO/STO superlattice from a set of 20 randomly-distorted $P1$ starting structures. Conventions as in Figure \ref{PTOmerge}. Red, blue and green arrows point to $Pc_{GSS}$, $Pc_{high}$  and $Cm_{GSS}$ states, respectively. The $Cm_{high}$ state does not appear in this set.}
 \end{figure}

As a complementary approach to investigating the energy surface for the 0\% case, 
we generated and relaxed twenty randomly-distorted starting structures\cite{note1}. The results are shown in Fig. \ref{randomP1}. 
Seven of the starting structures relax to the $Pc$ GSS, one relaxes to the $Cm$ GSS, and eight are at the same energy with a CT just barely larger than our threshold of 0.08\AA, indicating that they stay in $P1$ structure but are close to the $Cm$ or $Pc$ GSS, corresponding to intermediate directions of in-plane polarization. For the ground state, these results confirm the flatness of the energy surface suggested by our stacking method results. Four of the starting structures relax to the low-energy structure $Pc_{high}$. The low-energy state $Cm_{high}$ was missed in this process.

\begin{figure}
 \includegraphics[width=0.5\textwidth]{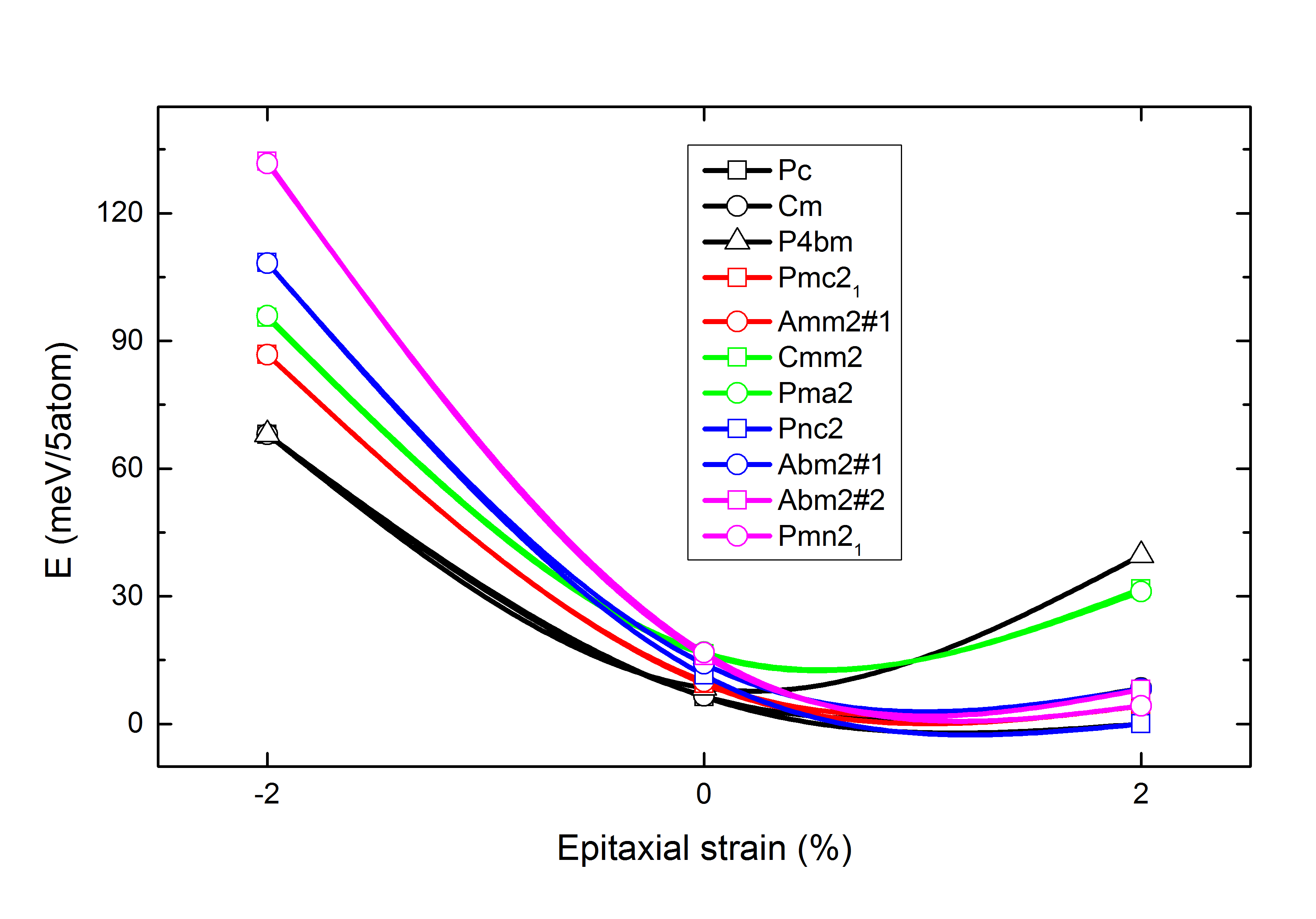}%
 \caption{\label{Connection}The total energies as functions of epitaxial strain for configurations. All the configurations fall into five distinct structures for -2\% strain, plotted in five colors. Different point types denote distinct space groups and show the revolution of them in epitaxial strain.}
 \end{figure}

Finally, it is instructive to put our results into the context of a more conventional ``energy curve" approach for constructing epitaxial phase sequences. This approach involves computation of the epitaxial strain dependence of the energy for relaxed structures based on selected distortions of the superlattice, plotting of the energies vs epitaxial strain, and analysis of the resulting curves to find the ground state structure and low energy structures at each strain. We performed additional calculations of the total energies of all the low-energy structures identified in our structure determination (Fig. \ref{PTOSTOdata}) for all three values of epitaxial strain, and present the resulting set of energy vs epitaxial strain curves in Fig. \ref{Connection}. 
At -2\% strain in Fig. \ref{Connection}, the stacking method identified the three lowest-energy configurations, $P4bm$, $P4/mbm$ and $P4mm$, the other two structures in this diagram being outside the 30 meV/5atoms energy range. 
For +2\% strain, the stacking method identified the four lowest-energy structures (shown as three points in Fig. \ref{Connection} due to the small energy differences), the other two structures being outside the energy window. 
In agreement with previous work\cite{Eric,Javier1}, at 0\% strain, many distinct structures are close in energy,
The energy curve approach is useful in understanding the evolution of symmetry-breaking distortions with strain and the resulting phase transitions, but involves energy computation for structures that are quite high in energy. In addition, the selection of distortions for the superlattice is generally not systematic. In comparison, our stacking method concentrates on the low-energy structures at each strain, and thus is more efficient for constructing the phase sequence and identifying the low-energy alternative structures at each strain.

\section{Discussion}


The example of PTO$_2$STO$_2$ demonstrates that the stacking method provides an efficient and systematic method for identifying the ground state and low-energy alternative structures for perovskite superlattices. The results at 0\% strain show that it is not enough to combine ground state structures of constituents, as has been assumed in some past studies, but that the ground state of the superlattice can be derived from alternative low-energy structures of the constituents. Further, the stacking method found ground-state and low-energy structures that had been missed by other methods. The approach is readily generalized to multicomponent perovskite superlattices and to superlattices based on other structure types. 

The method is particularly suitable for high-throughput studies of superlattices with constituents drawn from a specified set of pure compounds. Once the database of low-energy structures for the specified set of compounds is constructed, the generation of starting structures is rapid and automatic, and computational effort is focused on structures that are likely to be low in energy.

\section{Summary}

In summary, we have proposed a stacking method for the determination of the GSS and low-energy structures in perovskite superlattices. This method has been demonstrated in the 2:2 PTO/STO superlattice. For the range of epitaxial strain considered, our results for the GSS are consistent with previous work. For 0\% strain, this method 
highlights the previously unrecognized feature that the energy surface is ``flat'' near the ground state and hence the GSS is not well-defined. 
The method is double-checked by the random initial configuration method, and no structures with lower energy are found.
We have also shown the existence of two distinct structures with the same space group $Pc$ at 0\% strain, which could be difficult to identify using other methods.
This method allows for the efficient determination of the GSS and low-energy structures in general superlattice systems, paving the way for high-throughput studies of superlattices.

\section{acknowledgement}
We thank Sergey Artyukin, Premala Chandra, Kevin Garrity, Philippe Ghosez, Donald Hamann, Anil Kumar, Sebastian Reyes-Lillo, James Rondinelli, David Vanderbilt, Hongbin Zhang and Qibin Zhou for valuable discussions. 
This work is supported by NSF DMR-1334428, ONR N00014-11-1-0665, and ONR N00014-11-1-0666. 
First-principles calculations were done on the high performance computers in Rutgers University.

\bibliography {GSref}

\begin{thebibliography}{40}
\bibitem [{not()}]{note1}%
  \BibitemOpen
  \href@noop {} {}\bibinfo {note} {We have to know that in average it requires
  200 loops for the structural optimization if one relaxes the 2:2 PTO/STO from
  a random configuration, with the wall clock time 18 hours on 16 cores of
  x86-64 AMD Opteron. With the same computer resources, the computation cost
  for the ones with lowest symmetry, such as $Cm$ and $Pc$, in the stacking
  approach, is less than 4.2 hours for the same precision. The reduced
  computation time shows the importance of a reasonable initial configuration,
  and this advantage will be even amplified for large superlattices and
  high-throughput studies.}\BibitemShut {Stop}%
\end{thebibliography}
\end{document}